\documentstyle[11pt,epsf]{article}
\newcommand{\beq}{\begin{equation}}
\newcommand{\eeq}{\end{equation}}
\newcommand{\beqn}{\begin{eqnarray}}
\newcommand{\eeqn}{\end{eqnarray}}
\newcommand{\dd}{\mbox{d}}

\newcommand{\BSCCO}{$Bi_2Sr_2CaCu_2O_8$\,}
\newcommand{\rpr}{\vec{r}^{\prime}}
\newcommand{\QQ}{{\cal Q}}
\newcommand{\FF}{{\cal F}}

\newcommand{\tih}{\tilde{h}}
\newcommand{\tic}{\tilde{c}}
\newcommand{\ovl}[1]{\overline{#1}}
\advance \hoffset by -1.5cm
\advance \voffset by -2.0cm
\begin{document}
\vspace* {1.0cm}
\begin{center}
\Large{\bf Phase Diagram of a Classical Fluid in a Quenched Random
Potential}
\vskip 0.2in
\large{\bf Fabrice Thalmann$^{\dag,*}$, Chandan
Dasgupta$^{\dag}$ and Denis Feinberg$^{*}$}
\vskip 0.3in
$\dag$~Department of Physics\\ Indian Institute of
Science\\ Bangalore 560012, INDIA\\ and\\
Condensed Matter Theory Unit\\ Jawaharlal Nehru Centre for Advanced
Scientific Research\\
Bangalore 560064, INDIA.\\
\vskip 0.2in
$*$~Laboratoire d'Etudes des Propri\'et\'es Electroniques des Solides \\
Centre National de la Recherche Scientifique\\
BP166, 38042 Grenoble Cedex 9, FRANCE.\\
Laboratoire associ\'e \'a l'Universit\'e Joseph Fourier
\end{center}
\vskip 0.3in
\centerline {\bf \large ABSTRACT}
\vskip 0.2in

\noindent We consider the phase diagram of a classical fluid in the
presence of a random pinning potential of arbitrary strength.
Introducing replicas for averaging over the quenched disorder, we use
the hypernetted chain approximation to calculate the correlations in
the replicated liquid.  The freezing transition of the liquid into a
nearly crystalline state is studied using a density functional
approach, and the liquid-to-glass transition is studied using a
phenomenological replica symmetry breaking approach introduced by
M{\'e}zard and Parisi. The first-order liquid-to-crystal transition is
found to change to a continuous liquid-to-glass transition as the
strength of the disorder is increased above a threshold value.

\vskip 0.2in

\vskip 0.2in

The equilibrium phase diagram of a classical system of interacting
particles in the presence of quenched pinning potential is a subject of
much current interest, in view of understanding the behavior of
various systems such as magnetic bubble arrays~\cite{SeshWest}, Wigner
crystals of electrons~\cite{Andrei-et-al}, and flux lines in the mixed
phase of high-T$_c$ superconductors~\cite{BlaGesFeiLarVin}. In
particular, for layered type-II superconductors such as \BSCCO in a
magnetic field $H$ perpendicular to the layers, the ($T, H$) phase
diagram is especially interesting.  The flux lines in these materials
may be viewed as columns of interacting ``pancake'' vortices residing
on the layers, and the properties of the mixed phase may be described
in terms of the classical statistical mechanics of these point-like
objects. At low enough fields, a first-order melting transition
separates an ordered vortex lattice from a disordered ``vortex
liquid'' state~\cite{Zeldov-et-alii}. Existing theoretical 
studies~\cite{Nat,GiaLeD} suggest that weak point disorder only slightly
distorts the crystalline state, leading to the observed ``Bragg
glass'' phase with quasi-long-range translational
order~\cite{Cubitt-et-alii}. The freezing line of a layered system of
pancake vortices in the presence of weak point pinning has been
studied theoretically~\cite{MenDas}, assuming the solid phase to 
be a harmonic crystal. When the magnetic field is
increased, or equivalently with increasing
disorder~\cite{Khaykovitch-et-alii}, the experimental transition
becomes continuous~\cite{Zeldov-et-alii,Safar-et-al}, and the ``Bragg
glass'' is replaced by an amorphous state, which is believed to be a
vortex glass~\cite{F2-Huse,comment}.

This scenario could be a very general one, e.g. the first-order 
liquid-to-crystal transition in a three-dimensional (3d)
fluid may be driven by quenched disorder into a continuous
liquid-to-glass transition. In the present Letter, 
we consider, rather than vortices, a simpler and better
documented system, namely a fluid of hard spheres in a random pinning
potential of arbitrary strength. This is also a richer system
since it offers the possibility of an intrinsic, though metastable,
glassy phase, and the stabilization of this phase by quenched disorder
is of considerable interest.  We expect that our results could be
generalized and applied to the systems listed above. 
Using two ``mean-field''- type approaches based on the ``replicated liquid
formalism''~\cite{Pitard-et-al,MenDas,MezPar}, we obtain a phase
diagram in the density (~$\rho$) -- disorder (~$\delta$) plane which
shows crystalline, liquid and glassy phases. One expects that : i) the
transition between crystal and liquid phases remains first-order,
while the transition point itself is shifted with increasing disorder;
ii) from earlier work~\cite{MenDas} and arguments~\cite{ErtNel} based
on the Lindemann criterion, the liquid phase is favored by the
disorder. Our phase diagram is consistent with these expectations. We
also find that the first-order crystallization transition is replaced
by a continuous glass transition as the disorder strength is increased
above a threshold value.

We consider a fluid of monodisperse hard spheres in the presence of an
external, short-range, random pinning potential $\Phi(\vec{r})$.  In
the absence of disorder, the crystal becomes the stable phase at $\eta
= \eta_f \simeq 0.49$ ($\rho \simeq 0.94$) where the packing fraction
$\eta$ and the dimensionless density $\rho$ are related by
$\eta=\pi\rho/6$ (spheres of diameter 1).  If the hard-sphere fluid is
kept in a ``supercompressed'' state, then slow dynamics effects,
analogous to those occurring near the structural glass transition in
supercooled liquids, are observed~\cite{WoodAng}. Whether a true
thermodynamic glass transition occurs in this system at a packing
fraction lower than that for random close packing ($\eta_{rcp}\simeq
0.64$) is unclear~\cite{RinTor}. In analogy with experiments, one may
define the ``glass transition'' to occur at the point where the
characteristic relaxation time in the fluid exceeds a given value
$\tau_c$.  Then, the phase diagram of the pure hard-sphere fluid would
show a crystalline phase that is stable for $\eta > 0.49$, and at
$\eta=\eta_g$, just below $\eta_{rcp}$, a transition from the
(metastable) liquid to a (metastable) glassy phase.

The random potential $\Phi(\vec{r})$ is assumed to be gaussian with an
exponentially decreasing correlator ${\cal V}(\vec{r}-\rpr)= \ovl{
\Phi(\vec{r}) \Phi(\rpr)} $
\hbox{$=\Delta\cdot\exp[-(|\vec{r}-\rpr|^2)/\xi^2]$}; $\ovl{\Phi} = 0$,
($\ovl{\,\cdot\,}$ means averaging over the probability distribution
of $\Phi$), and a correlation length $\xi=0.25$. This choice implies
that local minima of $\Phi$ contain at most one sphere. The
parameter $\Delta$ measures the strength of the disorder. As soon as
an external potential is introduced, the temperature becomes a
relevant parameter~: the disorder may be considered perturbative if
$\delta \equiv \beta^2\Delta \ll 1;\; \beta=1/k_B T$, while $\delta
\gg 1$ corresponds to the strong pinning limit.

We make use of the ``replicated liquid formalism'', introduced for
studying fluids in porous media~\cite{Pitard-et-al}, flux-lattice
melting~\cite{MenDas}, and more recently, the structural glass
transition~\cite{MezPar}. The replica method is used for averaging
over the disorder and mapping the problem of an inhomogeneous liquid
in a random potential onto the problem of a homogeneous,
multicomponent mixture on which standard tools of liquid theory can be
applied. Before averaging over the disorder, the particles are coupled
to $\Phi(\vec{r})$ and interact via a hard-sphere repulsion.  After
introducing $n$ copies (``replicas'', labelled $a,b =1,\ldots, n$) of
the system and averaging over the gaussian disorder, one is left with
a mixture of $n$ components, interacting with $v_*(r) = -\beta {\cal
V}(r)$ plus a hard-sphere repulsion if the two particles belong to the
same replica, and with $v_0(r)=-\beta {\cal V}(r)$ otherwise.

The structure of the replicated liquid is expressed in terms of the pair
correlation functions $g_{ab}(r)=1+h_{ab}(r)$ between particles of
replicas $a$ and $b$. Direct correlations functions $c_{ab}(r)$ are
introduced, related to the $h_{ab}$ via the Ornstein-Zernike
relation, expressed in Fourier space as
\begin{equation}
  \forall a,b;\; \tih_{ab}(q) = \tic_{ab}(q) + \sum_d \tih_{ad}(q) \cdot
  \rho_d \cdot \tic_{db}(q).
  \label{eq:OZ}
\end{equation}
Following Refs~\cite{MenDas,MezPar}, we use the closure
scheme knows as the hypernetted chain (HNC) approximation in which the
$n \to 0$ limit can be taken analytically.
In our case, the HNC equations read~:
\begin{equation}
  g_{ab}(r)  = (1-\delta_{ab}+\delta_{ab}\cdot\Theta(r-1))\cdot
  \exp[\beta^2 {\cal V}(r) +h_{ab}(r) -c_{ab}(r)].
\label{eq:HNC}
\end{equation}
At relatively low values of the packing fraction, replica symmetry is a
natural assumption. Then, the set of functions $g_{ab}$ ($h_{ab}$ and
$c_{ab}$) reduces to $g_*$ ($h_*$ and $c_*$) if $a=b$ and $g_0$ ($h_0$
and $c_0$) if $a \neq b$, while all the densities $\rho_a$ are equal
to $\rho$. The function $g_*(r)$ looks very similar to the usual pair
correlation function of the pure system. In the absence of disorder,
$g_0(r)=1$, and it starts showing structure when $\delta$ is
increased. It exhibits peaks at the maxima of $g_*(r)$, plus an extra peak
at the origin induced by the attractive coupling between replicas.

The freezing into the nearly crystalline Bragg glass phase occurs in the
replica symmetric (RS) regime. According to the standard approach of
density functional theory~\cite{RamYou,MenDas}, the crystal manifests
itself through a (almost)
periodic modulation of the density $\rho_{cr}(\vec{r})$, obeying the
approximate self-consistency equation:
\begin{equation}
  \rho_{cr}(\vec{r}) = \rho\cdot \exp\left[ \int \dd \vec{r}\
C_{e}(\vec{r}-\rpr)
  \cdot (\rho_{cr}(\vec{r})-\rho)\right],
  \label{eq:DFT}
\end{equation}
where the function $C_{e}(\vec{r})$ stands for $\sum_b c_{ab}(\vec{r})
= c_* + (n-1)\cdot c_0(\vec{r}) = c_*(\vec{r}) - c_0(\vec{r})$ for $n
\to 0$ in the RS regime. The determination of the freezing
line in the presence of disorder then reduces to a standard density
functional calculation with a modified direct correlation function.

The glass transition in this system is studied using the
phenomenological approach of M\'ezard and Parisi (MP)~\cite{MezPar}
who found the occurrence of one-step replica symmetry breaking (RSB)
in equations (\ref{eq:OZ}),(\ref{eq:HNC}) in the absence of external
disorder and interpreted it as the emergence of a glassy phase. MP
consider a functional ${\cal F}$ of the pair correlation functions
$g_{ab}(r)$ which yields the set of
equations~(\ref{eq:OZ}),(\ref{eq:HNC}) upon functional differentiation
with respect to $g_{ab}$.  The one-step RSB solution consists, as
usual, in forming $n/m$ groups, each containing $m$
replicas~\cite{MezParVir}. Pair correlations $g_{ab}$ are set to be
$g_*$ for $a=b$, $g_1$ for $a \ne b$ with $a,b$ in the same group, and
$g_0$ if $a$ and $b$ are in different groups. Then, one looks for
solutions making the free energy ${\cal F}$ stationary with respect to
$g_*, g_1, g_0$ and $m$, with $0<m<1$ and $n=0$.  By analogy with
one-step RSB in mean-field spin-glass models with multispin
interactions, a ``dynamical transition'' density $\rho_{dyn}$ is
defined as the minimal density for which a RSB solution with $m=1$,
stationary with respect to $g_*,g_1,g_0$ but not with respect to $m$,
exists. The occurrence of a stationary solution with $m=1$ at a higher
value of $\rho$ signals a thermodynamic glass transition.

The approach of MP relies strongly on the assumption that in the
vicinity of the glass transition, the configuration space is split
into an exponentially large number of ``metastable
states''~\cite{MezParVir}. These metastable states come into existence
at the ``dynamical transition'' density $\rho_{dyn}$, and the
thermodynamic glass transition is expected to occur at the density where the
configurational entropy associated with the metastable states 
vanishes~\cite{Mon,KirThi}. Since the concept of metastable
states is itself a mean-field one borrowed from the study of
infinite-range spin-glass models, it is not clear whether this
description survives in real 3d fluids. Nevertheless, we believe that
both thermodynamics and dynamics of realistic 3d fluids near the glass
transition are dominated by long-lived
configurations~\cite{Dasgupta,SasDebSti}, which would be the 3d analog
of the metastable states.  Then, the method proposed by MP may be
regarded as a phenomenological way of finding when these long-lived
states start playing a significant role. We also note that results
very similar to those of MP have been obtained in a
calculation~\cite{CarFraPar} that uses a different method for locating
the glass transition in the pure hard-sphere system.

Due to the finite dimensional character of the system, the
``overlaps'' $g_1(r),g_0(r)$ have a spatial structure which accounts
for the short-range ordering of the particles in the metastable
states. Roughly speaking, $g_1(r)$ describes short-range correlations
of the average local density in a typical metastable state, and
$g_0(r)$ represents such correlations between different metastable
states [$g_0(r)=1$ in the absence of disorder].  We define the scalar
${\cal Q} = [ \int \dd r\ 4\pi r^2\ (g_1(r)-g_0(r))^2 ]^{1/2}$ as a
global order parameter.  At zero disorder, the RSB transition occurs
at $\rho_g \simeq 1.16$ and is discontinuous, with a jump in the value
of $\QQ$ at the transition~\cite{comment:B}.

We have solved the replicated liquid-state equations using the method
proposed by Zerah~\cite{Zer}, and a grid of 512 points.  Fig.~1(a)
presents our phase diagram. The thermodynamic glass transition found
in~\cite{MezPar} extends (thick line) from $\rho \simeq 1.16$,
$\delta=0$ to a ``tricritical'' point~T~: $\rho_t \simeq 1.11$,
$\delta_t \simeq 1.2$.  For $\delta<\delta_t$, $\QQ$ jumps
discontinuously at the transition, while for $\delta>\delta_t$ (thin
line), $\QQ$ grows continuously from zero as $\rho$ is increased
across the transition line.  The dashed line represents the freezing
line obtained from the density functional calculation. The liquid
phase is favored by the disorder and the freezing line crosses the
glass transition line at~C: $\rho_c \simeq 1.10$, $\delta_c \simeq
1.67$. The glass phase is also favored by the disorder until $\delta
\simeq 1.7$ where the liquid shows a re-entrance.  The upper part of
the freezing line lies in a region where the disorder is not weak
($\delta > 1$) and so, the stability of the Bragg glass phase is not
ensured. However, if we assume that the ordered phase remains stable
in this region, then extrapolation of the freezing line into the
glassy domain leads to the inset of Fig.~1(a), where only stable
phases are shown. As the density is increased at constant $\delta$,
the system undergoes a first-order transition to a nearly crystalline
state if $\delta<\delta_c$, and a continuous glass transition for
$\delta>\delta_c$. Thus, the phase diagram exhibits a multicritical
point where a line of continuous liquid-glass transition ends at a
line of first-order transitions representing a liquid-crystal
transition for $\delta<\delta_c$ and a glass-crystal transition for
$\delta>\delta_c$.

We now provide some of the technical details of our calculations.
The HNC free-energy per unit volume $V$ reads~:
\begin{eqnarray}
\frac{2\beta \FF}{nV}  & = & \rho^2\int\dd \vec{r}\ \left\lbrace
   g_*\, (\ln g_* -1+\beta v_*) + (m-1)\, g_1\, (\ln g_1 -1
   +\beta v_0) -m\, g_0\, (\ln g_0-1 +\beta v_0) \right\rbrace \nonumber \\
            &   & \hspace{-0.5cm} + \int \frac{\dd
   \vec{q}}{(2\pi)^3} \left\lbrace \rho \tih_* -\frac{\rho^2}{2}
   (\tih_*^2 +(m-1)\tih_1^2 -m\tih_0^2 ) + \frac{1-m}{m} \ln(1+
   \rho\tih_* -\rho\tih_1 ) \right. \nonumber\\
            &   & \hspace{-0.5cm} \left. -\frac{1}{m} \ln(1
   +\rho\tih_* +(m-1)\,\rho\tih_1  -m\,\rho\tih_0)
   -\frac{\rho\tih_0} { [1 +\rho\tih_* + (m-1)\,\rho\tih_1
   -m\,\rho\tih_0]} \right\rbrace.
\end{eqnarray}
In the $n\to 0$ limit, a physically stable solution must be a minimum
with respect to variations of $g_*$, but a maximum with respect to
$g_1,g_0$. As $m \to 1$, the equations for $g_1$ decouple from those
of $g_*$ and $g_0$, and one can get simultaneously the RSB solution
$(g_*,g_1 \ne g_0)$ and the RS one $(g_*,g_1=g_0)$. Both stable RS and
RSB solutions exist in the domain bounded by the lines (IN,DT,
$\delta=0$) in Fig.~1(b). The thermodynamic glass
transition (line TGT) occurs when $\partial \FF/\partial m|_{m=1}=0$,
\textit{i.e.} when the second term $\Phi'[g_*,g_1,g_0]$ of the
expansion $\FF = \Phi[g_*,g_0] + (m-1) \cdot \Phi' +
{\cal O}((m-1)^2)\ldots$ vanishes (this is our practical
criterion for determining the location of the line TGT).

Examples of RS and RSB solutions on two sides of the line TGT at
$\delta=0.3$ are shown in Fig.~2(a). These plots illustrate
the discontinuity in $\QQ$ across the line TGT. This discontinuity decreases
with increasing $\delta$, and
the transition  eventually becomes continuous for $\delta \geq
\delta_t$ on the line CT~(continuous transition).  Fig.~2(b)
presents the RSB solution next to the line
CT at $\delta=1.8$. The functions $g_1$ and $g_0$ look very similar,
indicating that $\QQ$
vanishes at the transition, as shown in the inset. The transition at
$\delta > \delta_t$ occurs via a bifurcation mechanism. This can
be understood by expanding $\FF$ to second order in $\Delta
g_*, \Delta g_1, \Delta g_0$ around the RS solution.  Defining
$\Delta k_1 = \sqrt{m(1-m)} (\Delta h_1 -\Delta h_0)$, $\Delta k_0 =
(1-m) \Delta h_1 + m \Delta h_0$, and $A= 1 +\rho \tih_*-\rho\tih_0$,
we get~:
\begin{equation}
\frac{\beta \Delta\FF}{nV}  =   {\cal J}(\Delta h_*, \Delta k_0)
                - \rho^2/4 \left\lbrace  \int \dd\vec{r}\
                 (g_0^{-1}-1) \Delta k_1^2 + \int \dd
                 \vec{q}/(2\pi)^3\ A^{-2} \Delta \tilde{k}_1^2
                 \right\rbrace,
\label{eq:hessian}
\end{equation}
where ${\cal J}$ is a $m$-independent quadratic form.  We have
diagonalized a discrete version of the quadratic form enclosed in the
braces, and found that its lowest eigenvalue $\lambda_{min}$ changes
its sign, from positive (in the liquid phase) to negative (in the
glass phase), just when $\QQ$ vanishes (inset of
Fig.~2(b)). The RS solution turns unstable as
$\lambda_{min}$ becomes negative, providing a precise determination of
the line CT. This line
admits a continuation for $\delta < \delta_t$, labelled IN
(instability line) in Fig.~1(b), which does
not seem to have any physical meaning. A similar computation around
the RSB solution enables us to determine precisely the bifurcation
point at which the RSB solution, $m=1$, appears. This calculation yields the
``dynamical transition'' line DT.  As the
transition becomes continuous near $\delta \sim \delta_t$, finding
numerically the line TGT becomes very difficult because the $\Phi'$
variations become vanishingly small. However, the fact that the lines
CT, DT and IN converge to the same point T leads us to the
conclusion that the TGT line also ends at T, as depicted in
Fig.~1(a). 

The pure hard-sphere fluid freezes into a fcc lattice. We use the
Ramakrishnan-Yussouff method~\cite{RamYou} for expressing the density
$\rho_{cr}(\vec{r})$ in terms of three ``order parameters'' --
the Fourier components, $\mu_{1,1,1},\mu_{3,1,1}$, of the
density for two sets ($(1,1,1)$ and $(3,1,1)$) of reciprocal lattice
vectors, and $\eta$, the fractional change in the average density:
\begin{equation}
  \ln\frac{\rho_{cr}}{\rho} = \left[ \eta \rho\ \tilde{C}_{e}(0) +
  \mu_{1,1,1} \sum_{(1,1,1)} \tilde{C}_{e}(|K_{1,1,1}|)
  e^{i\vec{K}_{1,1,1}\cdot \vec{r}} + \mu_{3,1,1} \sum_{(3,1,1)}
  \tilde{C}_{e}(|K_{3,1,1}|) e^{i\vec{K}_{3,1,1}\cdot \vec{r}} \right]
\end{equation}
We believe that this parametrization is justified in the
presence of weak disorder because the nearly crystalline Bragg glass
phase exhibits well-defined peaks in its structure factor. The HNC direct
correlation function is used as input, and the coexistence line is
obtained from the usual thermodynamic criterion~\cite{RamYou}:
\begin{equation}
  \int \dd \vec{r}\ \left(\rho_{cr}\cdot
  \ln(\frac{\rho_{cr}}{\rho}) -\rho_{cr}+\rho
  \right) -\frac{1}{2}
  \int \dd \vec{r}\! \int \dd \vec{r}^{\ '}\ C_{e}(\vec{r}-\vec{r}^{\
  '}) \cdot (\rho_{cr}'-\rho)\cdot(\rho_{cr}-\rho)
  = 0.
\end{equation}
At $\delta=0$, we get $\rho_f=0.93$, with $\eta \simeq 15\%$.  As $\delta$
is increased, $\rho_f$ increases (see line FL in Fig.~1(b))
and $\eta$ decreases, reaching a value of about 6\% near the point C.

To summarize, we have used a combination of the replica method, liquid
theory and density functional theory to obtain the phase diagram of a
simple classical fluid in a random pinning potential. Our calculations
quantify the effects of the disorder on the crystallization transition
of the pure fluid and its glass transition in the metastable
``supercompressed'' regime. We find that
the first-order crystallization transition of the pure fluid changes
to a continuous glass transition as the strength of the disorder is
increased above a critical value. The phase diagram we have obtained
looks qualitatively similar to that of layered type-II
superconductors if, instead of
increasing the density $\rho$, one decreases the temperature, and if one
replaces the disorder strength $\delta$ by the magnetic
field $H$. Since our calculations are
mean-field in nature, they do not provide a conclusive answer to the
question of whether a thermodynamic glass phase exists in 3d randomly
pinned classical systems. Further investigations of this issue and
extensions of our calculation to layered superconductors and other
physical systems would be very interesting.

We thank M. M\'ezard, G. Parisi, J.P. Hansen and G. Menon for
stimulating discussions. This work was supported in part by the
International Program PICS No. 482 between CNRS and Jawaharlal Nehru
Centre for Advanced Scientific Research (JNC). Two of the authors
(F.T. and D.F.) gratefully acknowledge support and hospitality from
Indian Institute of Science and JNC.


\newpage
\centerline{\Large {Figure captions}}
\vspace*{0.5cm}

\noindent {\bf Fig.~1}: (a) The phase diagram in the
density~$\rho$ -- disorder~$\delta$
  plane. T is a tricritical point where the nature of the
  glass transition changes from first-order to continuous, and C is a
  critical endpoint where the continuous glass transition line meets the
  first-order freezing line. The inset depicts the phase diagram in
  which only the
  thermodynamically stable phases are shown.

\noindent  (b) Various transition lines (see text for details) in the
  $\rho-\delta$ plane. TGT: first-order glass transition
  for weak disorder;
  CT: continuous glass transition for strong disorder;
  DT: dynamical transition; IN: instability of the RS solution;
  FL: freezing line obtained from density functional theory.

\vspace*{0.5cm}

  \noindent {\bf Fig.~2}:
  (a): HNC pair correlation functions near TGT, at
  $\delta=0.3$. Thick and dashed lines stand respectively for $g_1$
  and $g_0$ at $\rho =1.15$, slightly on the glass side, while the
  ``diamond curve'' is the replica symmetric solution at $\rho=1.14$,
  on the liquid side. Note the jump of $\QQ$ at the
  transition. Inset~: $g_*,g_1,g_0$ at $\delta=0.3,\rho=1.15$.

\noindent (b): Pair
  correlation functions near the continuous transition CT, at
  $\delta=1.8$. Inset~: The ``order parameter'' $\QQ$ (see text) which
  vanishes at the transition point. Also shown is $\lambda$, the lowest
  eigenvalue of the
  quadratic form enclosed in braces~ in Eq.(\ref{eq:hessian}), which
  changes its sign at the same point.
\newpage
\begin{figure}[htbp]
  \begin{center}
    \leavevmode
    \epsfysize=10.0cm\epsfbox{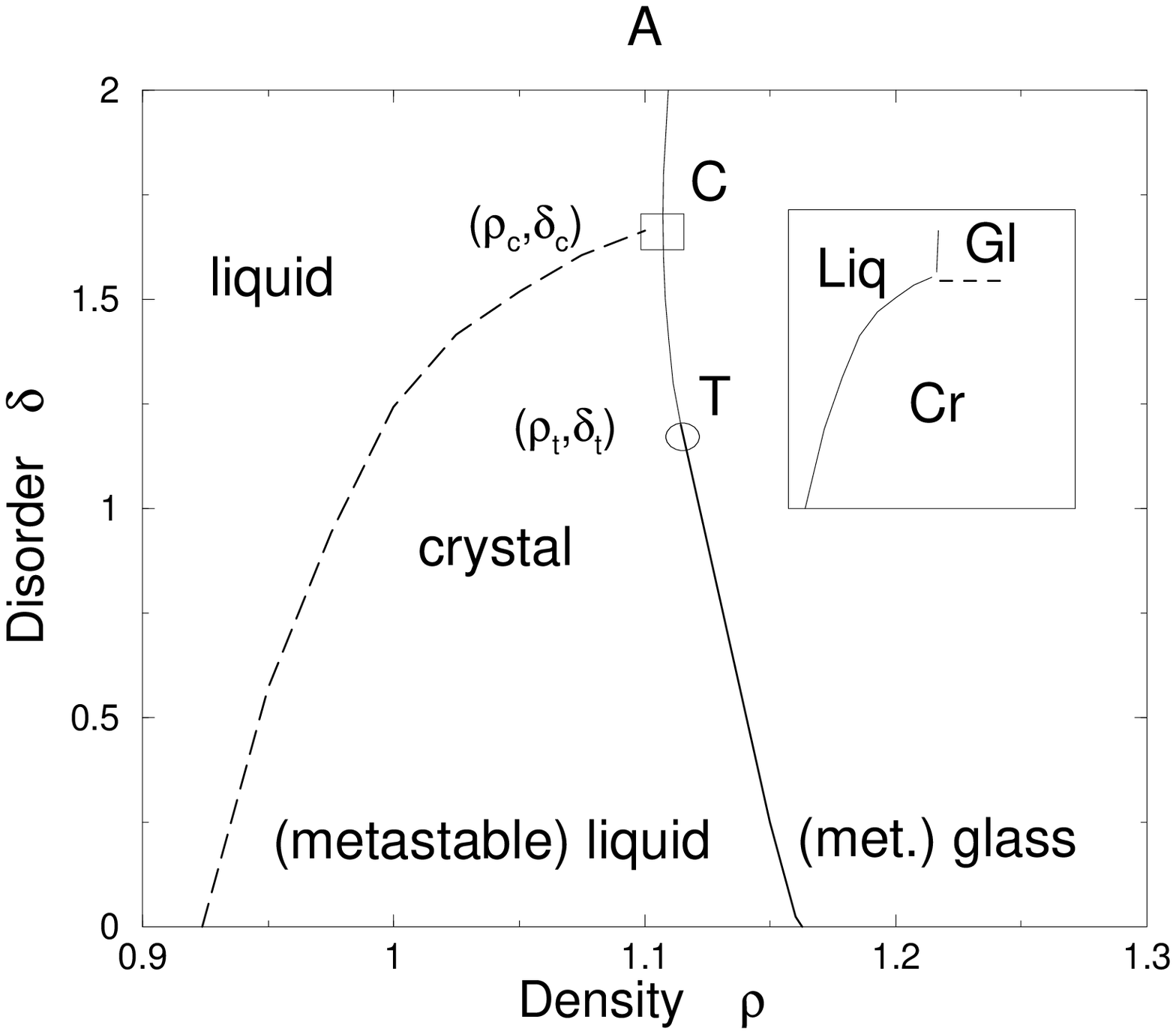}
  \end{center}
\centerline{(a)}
\begin{center}
    \epsfysize=10.0cm\epsfbox{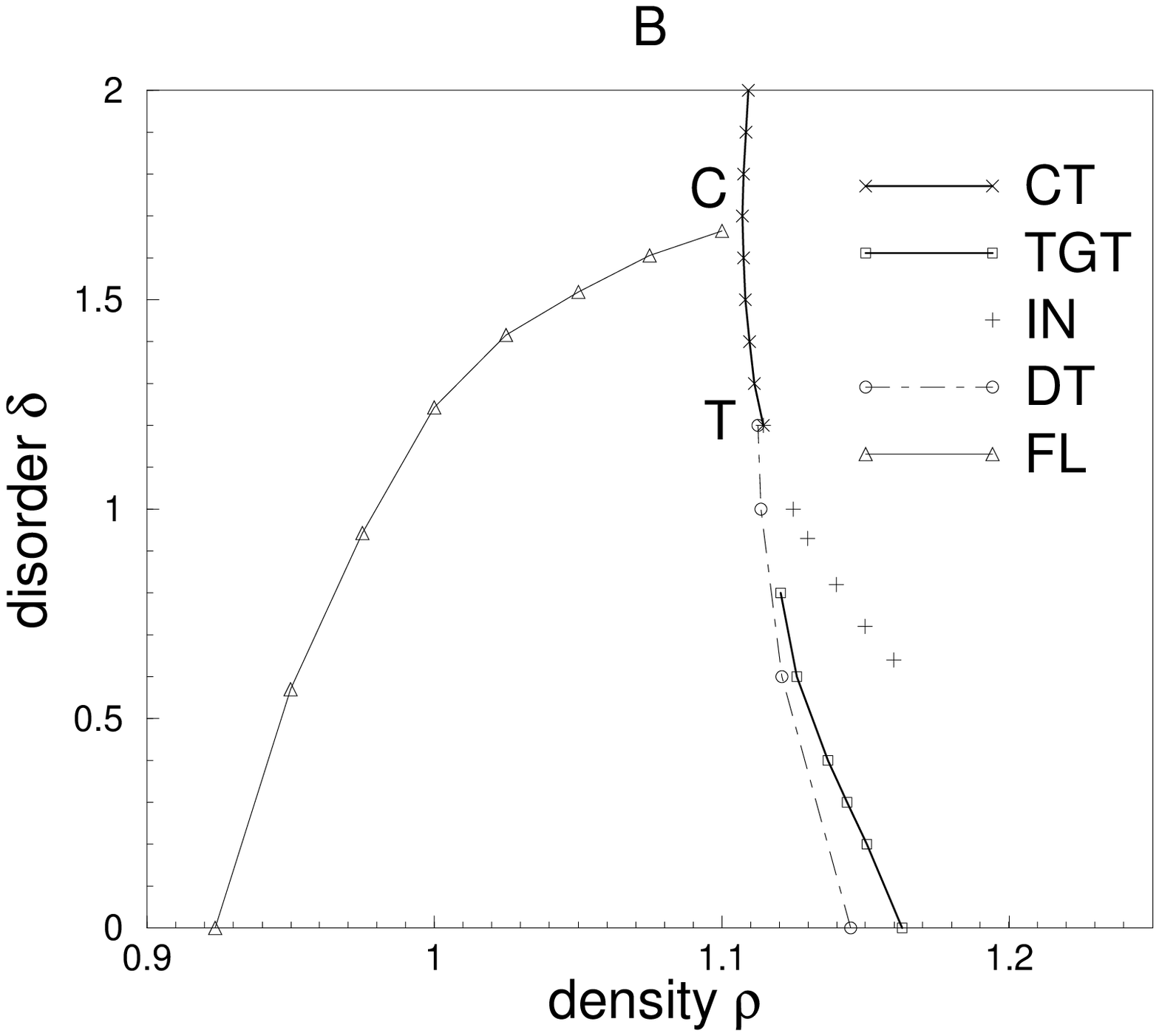}
  \end{center}
\centerline{(b)}
\vspace*{0.3cm}
\centerline{\Large {Fig.~1}}
\label{fig:one}
\end{figure}

\newpage
\begin{figure}[htbp]
  \begin{center}
    \leavevmode \epsfysize=10.0cm \epsfbox{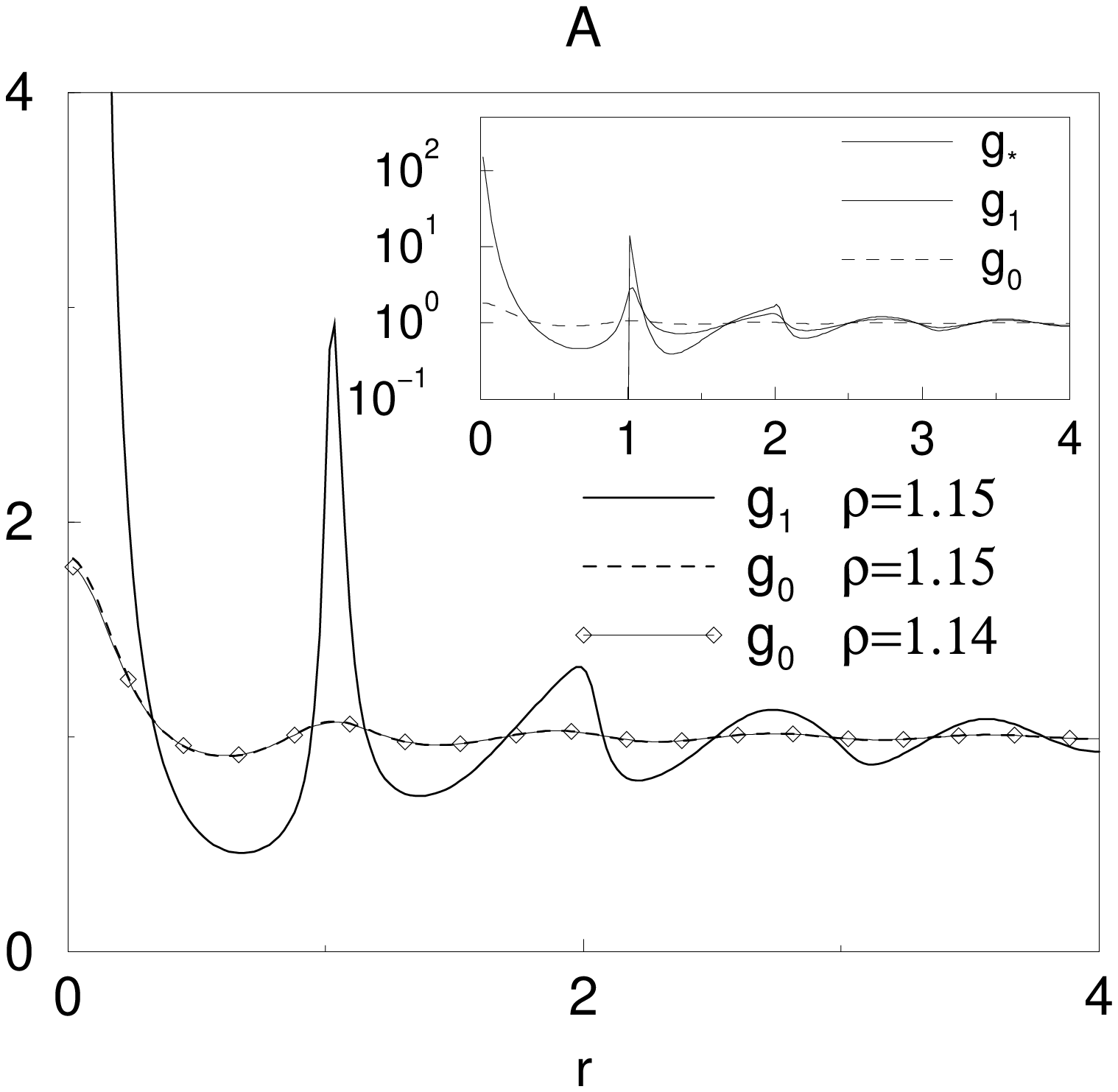}
\end{center}
\centerline{(a)}
\begin{center}
    \epsfysize=10.0cm \epsfbox{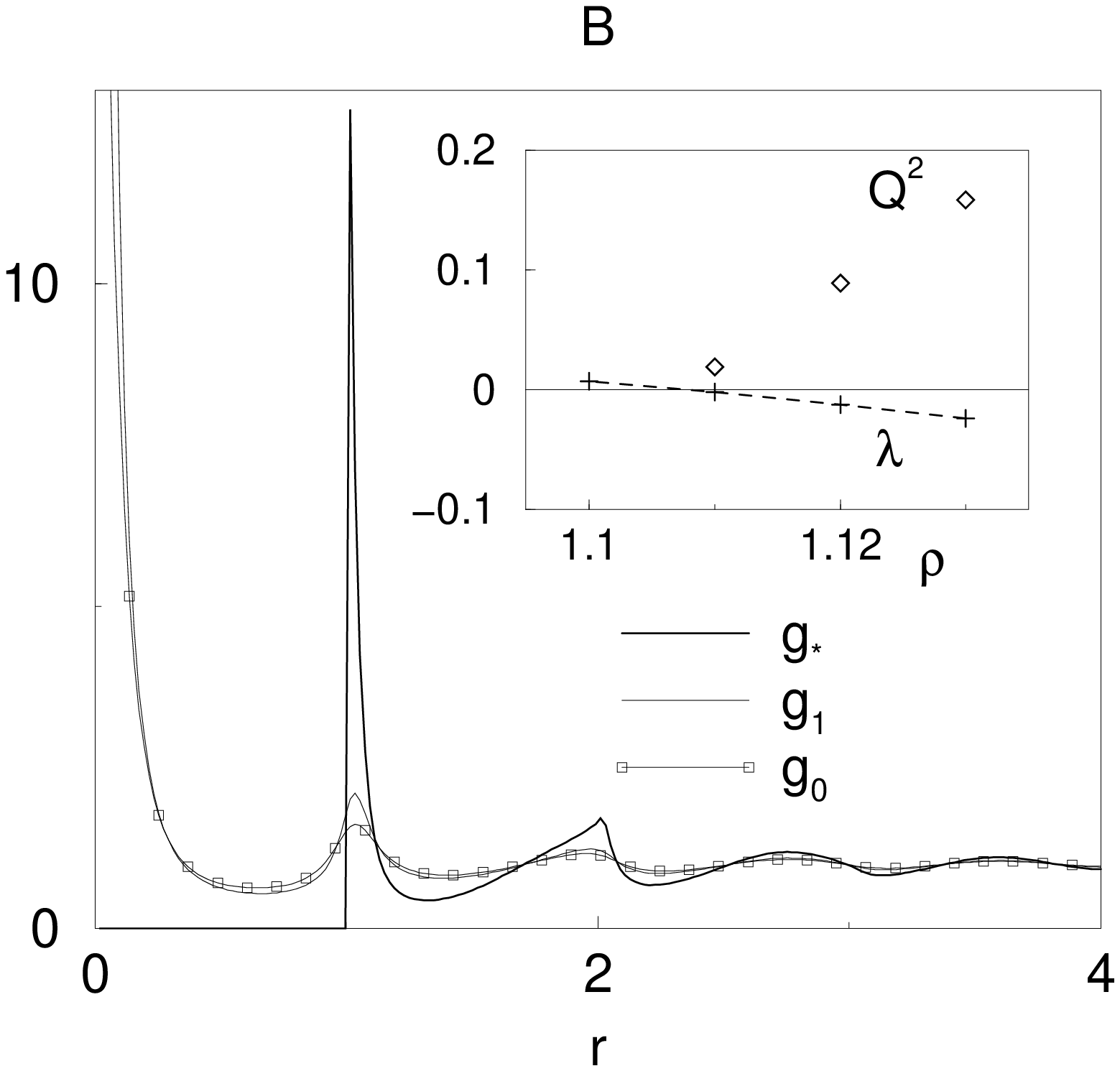}
  \end{center}
\centerline{(b)}
\vspace*{0.3cm}
\centerline{\Large{Fig.~2}}
\label{fig:two}
\end{figure}


\newpage
\begin{thebibliography}{99}\frenchspacing
\bibitem{SeshWest} SESHADRI R. and WESTERVELT R. A., {\it Phys. Rev.
B}, {\bf 46} (1992) 5150.
\bibitem{Andrei-et-al} ANDREI E. Y. {\it et al.}, {\it Phys. Rev.
Lett.}, {\bf 60} (1988) 2765.
\bibitem{BlaGesFeiLarVin} BLATTER G. {\it et al.}, {\it  Rev. Mod. Phys.},
{\bf 66} (1994) 1125.
\bibitem{Zeldov-et-alii} ZELDOV E. {\it et al.}, {\it Nature},
{\bf 375} (1995) 373.
\bibitem{Nat} NATTERMANN T., {\it Phys. Rev. Lett.},
{\bf 64} (1990) 2454.
\bibitem{GiaLeD} GIAMARCHI T. and LE DOUSSAL P., {\it Phys. Rev. B},
{\bf 52} (1995) 1242.
\bibitem{Cubitt-et-alii} CUBITT R. {\it et al.}, {\it Nature},
{\bf 365} (1993) 407.
\bibitem{MenDas} MENON G. and DASGUPTA C., {\it Phys. Rev. Lett.},
{\bf 73} (1994) 1023.
\bibitem{Khaykovitch-et-alii} KHAYKOVITCH B. {\it et al.}, {\it Phys. Rev. B},
{\bf 56} (1997) R517.
\bibitem{Safar-et-al} SAFAR H. {\it et al.}, {\it Phys. Rev.
Lett.}, {\bf 70} (1993) 3800.
\bibitem{F2-Huse} FISHER D. S., FISHER M. P. A. and HUSE D. A., {\it
Phys. Rev. B}, {\bf 43} (1990) 130.
\bibitem{comment} The question of whether a true thermodynamic glass
phase exists in superconductors with random point pinning is
controversial - see, e.g., BOKIL H. S. and YOUNG A. P., {\it Phys. Rev.
Lett.}, {\bf 74} (1995) 3021.
\bibitem{Pitard-et-al} PITARD E. et al {\it Phys. Rev. Lett.}, {\bf
74} (1995) 4361. 
\bibitem{MezPar} M{\'E}ZARD M. and PARISI G., {\it J. Phys. A },
{\bf 29} (1996) 6515.
\bibitem{ErtNel} ERTA\c{S} D. and NELSON D. R., {\it Physica}, {\bf 271C}
(1996) 79.
\bibitem{WoodAng} WOODCOCK, L. V. and ANGELL, C. A., {\it Phys. Rev.
Lett.}, {\bf 47} (1981) 1129.
\bibitem{RinTor} RINTOUL M. and TORQUATO S., {\it Phys. Rev. Lett.},
{\bf 77} (1996) 4198.
\bibitem{RamYou} RAMAKRISHNAN T. V. and YUSSOUFF M., {\it Phys. Rev. B},
{\bf 19} (1979) 2775.
\bibitem{MezParVir} M{\'E}ZARD M., PARISI G. and VIRASORO M. A.
{\it Spin glass theory and beyond.}, World Scientific, 1987.
\bibitem{Mon} MONASSON R. {\it Phys. Rev. Lett.}, {\bf 75} (1995) 1170. 
\bibitem{KirThi} KIRKPATRICK T. R. and THIRUMALAI D. {Phys. Rev. B},
{\bf 36} (1987) 5388.
\bibitem{CarFraPar} CARDENAS M., FRANZ S. and PARISI G.,
{\it J. Phys. A}, {\bf 31} (1998) L163.
\bibitem{Dasgupta} DASGUPTA C., {\it Europhys. Lett.},
{\bf 20} (1992) 131.
\bibitem{SasDebSti} SASTRY S., DEBENEDETTI P. G. and STILLINGER F. H.,
{\it Nature}, {\bf 393} (1998) 554.
\bibitem{comment:B} The value given in~\cite{MezPar} is
{$\rho_g=1.19$}. The discrepancy comes from a different discretization
procedure. Given a solution on a real-space grid with spacing {$\Delta
r$}, and a hard-sphere diameter {$d$}, there is an uncertainty
{$\Delta \eta/\eta = 3\Delta r/d$} in the value of the associated
packing fraction, which accounts for the difference between the two
numerical results.
\bibitem{Zer} ZERAH G., {\it J. Comput. Phys}, {\bf 61}
(1985) 280.
\end{thebibliography}
\end{document}